\documentclass[aps,prd,showpacs,twocolumn,amsmath]{revtex4-1}
\usepackage{subfigure}
\usepackage{graphicx}
\usepackage{dcolumn}
\usepackage{bm}
\usepackage{float}
\usepackage{xcolor}
\usepackage{epstopdf}
\definecolor{aa}{RGB}{0,0,139}
\usepackage[bookmarksnumbered=true,colorlinks,urlcolor=aa,linkcolor=blue,anchorcolor=aa,citecolor=aa]{hyperref}
\usepackage[mathlines]{lineno}
\usepackage{multirow}
\usepackage{enumerate}

\usepackage{amsmath}

\newcommand{\EE}{e^+e^-}
\newcommand{\ep}{\eta\phi}
\newcommand{\ISR}{\EE \rightarrow \phi\gamma_{\rm ISR}}
\newcommand{\kk}{K^+ K^-}
\newcommand{\GG}{\gamma\gamma}
\newcommand{\jpsi}{J/\psi}
\begin{document}
\title{Study of $\EE \to \ep$ at center-of-mass energies from 3.773 to 4.600~GeV }

\author{M.~Ablikim$^{1}$, M.~N.~Achasov$^{13,b}$, P.~Adlarson$^{75}$, X.~C.~Ai$^{81}$, R.~Aliberti$^{36}$, A.~Amoroso$^{74A,74C}$, M.~R.~An$^{40}$, Q.~An$^{71,58}$, Y.~Bai$^{57}$, O.~Bakina$^{37}$
% ...
\\
\vspace{0.2cm}
(BESIII Collaboration)\\
\vspace{0.2cm} {\it
$^{1}$ Institute of High Energy Physics, Beijing 100049, People's Republic of China\\
% ...
$^{13}$ G.I. Budker Institute of Nuclear Physics SB RAS (BINP), Novosibirsk 630090, Russia\\
% ...
$^{36}$ Johannes Gutenberg University of Mainz, Johann-Joachim-Becher-Weg 45, D-55099 Mainz, Germany\\
$^{37}$ Joint Institute for Nuclear Research, 141980 Dubna, Moscow region, Russia\\
% ...
$^{40}$ Liaoning Normal University, Dalian 116029, People's Republic of China\\
% ...
$^{57}$ Southeast University, Nanjing 211100, People's Republic of China\\
$^{58}$ State Key Laboratory of Particle Detection and Electronics, Beijing 100049, Hefei 230026, People's Republic of China\\
% ...
$^{71}$ University of Science and Technology of China, Hefei 230026, People's Republic of China\\
% ...
$^{74}$ University of Turin and INFN, (A)University of Turin, I-10125, Turin, Italy; (B)University of Eastern Piedmont, I-15121, Alessandria, Italy; (C)INFN, I-10125, Turin, Italy\\
$^{75}$ Uppsala University, Box 516, SE-75120 Uppsala, Sweden\\
% ...
$^{81}$ Zhengzhou University, Zhengzhou 450001, People's Republic of China\\
\vspace{0.2cm}
% ...
$^{b}$ Also at the Novosibirsk State University, Novosibirsk, 630090, Russia\\
% ...
}}

\vspace{0.4cm}
\date{\today}

\begin{abstract}
We present a study of the process $\EE \to \ep$ using data samples collected with the BESIII detector corresponding to an integrated luminosity of 15.03~fb$^{-1}$ at 23 center-of-mass energies from 3.773 to 4.600~GeV.
The Born cross sections are measured at each energy and a coherent fit to cross-section lineshape is performed using a Breit-Wigner parametrization to search for charmonium-like vector states. No significant signals of the $Y(4230)$ and $Y(4360)$ resonances are observed.
\end{abstract}

\maketitle

\section{Introduction}
Since the $\chi_{c1}(3872)$ was discovered in 2003 \cite{Choi:2003ue}, a series of charmonium-like states have been reported in experiments at various laboratories using complementary probes, such as $\EE$ annihilations, $pp$ collisions, and via $B$ meson decays~\cite{Olsen:2017bmm,Brambilla:2019esw,Belle-II:2018jsg}. These new states exhibit some exotic properties, which are unexpected in the conventional charmonium spectrum. Particularly, precision data of the vector-meson $Y$ states in the charmonium-mass region have drawn attention in the research community.
The first $Y$ state, $Y$(4260), was observed by the BaBar Collaboration using the initial-state radiation (ISR) process $\EE \rightarrow \gamma_{\rm ISR}J/\psi\pi^{+}\pi^{-}$~\cite{BaBar:2005hhc}, and then confirmed by CLEO-c, Belle and BESIII~\cite{CLEO:2006tct,Belle:2007dxy,BESIII:2016bnd}. Furthermore, evidence for transitions from the $Y$(4260) to other charmonium-like states, such as the $\chi_{c1}$(3872) and $Z_c$(3900), have been reported~\cite{BESIII:2019qvy,BESIII:2020oph}. Beside $Y$(4260), $Y$(4360) (denoted as $\psi(4360)$ in Particle Data Group (PDG)~\cite{ParticleDataGroup:2022pth}) were observed in the $J/\psi\pi^{+}\pi^{-}$, $\psi(3686)\pi^{+}\pi^{-}$~\cite{BESIII:2021njb}, $h_c\pi^+ \pi^-$~\cite{BESIII:2016adj} and $J/\psi\eta$~\cite{BESIII:2020bgb} final states. Subsequently, further measurements by the BESIII confirmed the existence of the $Y$(4260) in other processes, including $J/\psi\pi^{0}\pi^{0}$, $J/\psi K^{+}K^{-}$~\cite{CLEO:2006ike}, $\omega \chi_{c0}$~\cite{BESIII:2019gjc}, $J/\psi\eta'$~\cite{BESIII:2019nmu},
and $\pi^+D^0D^{*-}$~\cite{BESIII:2018iea}. The new measurements by BESIII resulted in a downward shift of the $Y(4260)$ mass, which nowadays is referred to as the $Y(4230)$ (denoted as $\psi(4230)$ in PDG~\cite{ParticleDataGroup:2022pth}).

In spite of the major experimental and theoretical progress, the internal structure of these $Y$ states remains a mystery with controversial interpretations~\cite{Brambilla:2010cs}. Many alternative models have been proposed to interpret their nature with scenarios including conventional charmonium, tetraquarks, hadronic molecules, and hybrids~\cite{Guo:2017jvc,Chen:2016qju,Karliner:2017qhf,Berwein:2015vca,Guo:2013sya,Cleven:2013mka,Cleven:2013mkaa,Giron:2020fvd}.
To provide an unambiguous description of the internal structure of the observed $Y$ states and to conclude on their nature, further experimental information is highly desirable. In particular, searches for new decay modes will provide more information on their decay properties and, thereby, may shed light on the underlying production mechanisms.
Furthermore, with masses  between 4.0 and 4.6 GeV above the light unflavored meson thresholds, both $Y$ and excited $\psi$ states should couple to light unflavored final states, and many studies have been performed to measure the cross sections of two-body final states~\cite{BESIII:2017qkh,BESIII:2021yam,BESIII:2022tjc}. In such final states, new exotic particles and new decay modes of known $Y$ and excited $\psi$ states can be searched for.

In this paper, we report the measurements of Born cross sections of $\EE \rightarrow \ep$ at 23 center-of-mass (cm) energies from 3.773 to 4.600 GeV using data samples corresponding to an integrated luminosity of 15.03~fb$^{-1}$. An energy-dependent fit is performed to search for possible signals of the $Y(4230)$ and $Y(4360)$ states. These measurements are complementary to a recent BESIII study of the same final state performed at lower cm energies (2.0--3.08 GeV) in the vicinity of the $\phi(2170)$~\cite{BESIII:2021bjn}.

\section{BESIII detector and Monte Carlo simulation}
The BESIII detector~\cite{BESIII:2009fln} records symmetric $\EE$ collisions provided by the BEPCII storage ring~\cite{BEPCII}, which operates in the cm energy $\sqrt{s}$ range from 2.0 to 4.95 GeV, with a peak luminosity of $10^{33}\;\text{cm}^{-2}\text{s}^{-1}$ achieved at $\sqrt{s} = 3.773\;\text{GeV}$. The cylindrical core of the BESIII detector covers 93\% of the full solid angle and consists of a helium-based multilayer drift chamber (MDC), a plastic scintillator time-of-flight system (TOF), and a CsI(Tl) electromagnetic calorimeter (EMC), which are all enclosed in a superconducting solenoidal magnet providing a 1.0 T magnetic field~\cite{Huang:2022wuo}. The solenoid is supported by an octagonal flux-return yoke with resistive plate counter based muon identification modules interleaved with steel. The charged-particle momentum resolution at 1 GeV/$c$ is 0.5\%, and the resolution of the specific ionization energy loss in the MDC, d$E$/d$x$, is 6\% for electrons from Bhabha scattering. The EMC measures photon energies with a resolution of 2.5\% (5\%) at 1 GeV in the barrel (end cap) region. The time resolution in the TOF barrel region is 68 ps, while that in the end cap region is 110 ps. The end cap TOF system was upgraded in 2015 using multi-gap resistive plate chamber technology, providing a time resolution of 60 ps~\cite{BESIII:27a}.

The experimental data sets used in this analysis are listed in Table~\ref{tabB}. Data samples corresponding to an integrated luminosity of 8.05~fb$^{-1}$ were collected after the upgrade of the end cap TOF system.
Simulated data samples produced with a {\sc
geant4}-based~\cite{GEANT4:2002zbu} Monte Carlo (MC) package, which
includes the geometric description of the BESIII detector and the
detector response, are used to determine detection efficiencies
and to estimate backgrounds.
The signal MC samples of $\EE \rightarrow \ep$ at each energy point are simulated with the \textrm{ConExc} generator~\cite{Ping:2013jka}.
An inclusive MC sample equivalent to an integrated luminosity of 500~pb$^{-1}$ data set at $\sqrt{s}=4.258$ GeV is used to study the background. The inclusive MC sample includes
the production of open charm processes, the ISR production of vector charmonium(-like) states, and the continuum processes incorporated in {\sc kkmc}~\cite{ref:kkmc}.
All particle decays are modeled with {\sc evtgen}~\cite{ref:evtgen} using branching fractions either taken from the Particle Data Group~\cite{ParticleDataGroup:2022pth}, when available,
or otherwise estimated with {\sc lundcharm}~\cite{ref:lundcharm}.
Final state radiation~(FSR) from charged final state particles is incorporated using the {\sc photos} package~\cite{photos}.

\section{Event selection and background analysis}

To select the candidate events of the process $e^{+}e^{-}\rightarrow\eta\phi$, with $\phi\to K^{+}K^{-}$ and $\eta\to\gamma\gamma$, the following event selection criteria are applied to both data and MC samples.

The number of charged tracks is required to be two with opposite charges.
Charged tracks detected in the MDC are required to be within a polar angle ($\theta$) range of $|\rm{cos\theta}|<0.93$, where $\theta$ is defined with respect to the $z$-axis,
which is the symmetry axis of the MDC. For each track, the distance of closest approach to the interaction point (IP) must be less than 10\,cm along the $z$-axis, $|V_{z}|$,
and less than 1\,cm in the transverse plane, $|V_{xy}|$.
Particle identification~(PID) for charged tracks combines measurements of the energy deposited in the MDC~(d$E$/d$x$) and the flight time in the TOF to form likelihoods $\mathcal{L}(h)~(h=p,K,\pi)$ for each hadron $h$ hypothesis.
Two tracks are identified as kaons when the kaon hypothesis has the greatest likelihood ($\mathcal{L}(K)>\mathcal{L}(\pi)$ and $\mathcal{L}(K)>\mathcal{L}(p)$).

Photon candidates are identified using showers in the EMC.  The deposited energy of each shower must be more than 25~MeV in the barrel region ($|\cos \theta|< 0.80$) and more than 50~MeV in the end cap region ($0.86 <|\cos \theta|< 0.92$).  To suppress electronic noise and showers unrelated to the event, the difference between the EMC time and the event start time is required to be within
[0, 700]\,ns. The number of photon candidates should be 2 or greater.

After performing a primary vertex fit to two charged tracks, a four-constraint (4C) kinematic fit is applied under the hypothesis $\EE\to\kk\GG$ constraining the toal four-momentum of the final state particles to match that of the initial $\EE$ system. For events with more than two good photon candidates, the two photons are picked out from all combinations with the $K^+K^-$ pair having the minimum $\chi^2$ value of the kinematic fit. Candidates with $\chi^2<100$ are retained for further analysis.

\begin{figure}[h]
  \centering
  \includegraphics[width= 8.9cm,height=6.0cm]{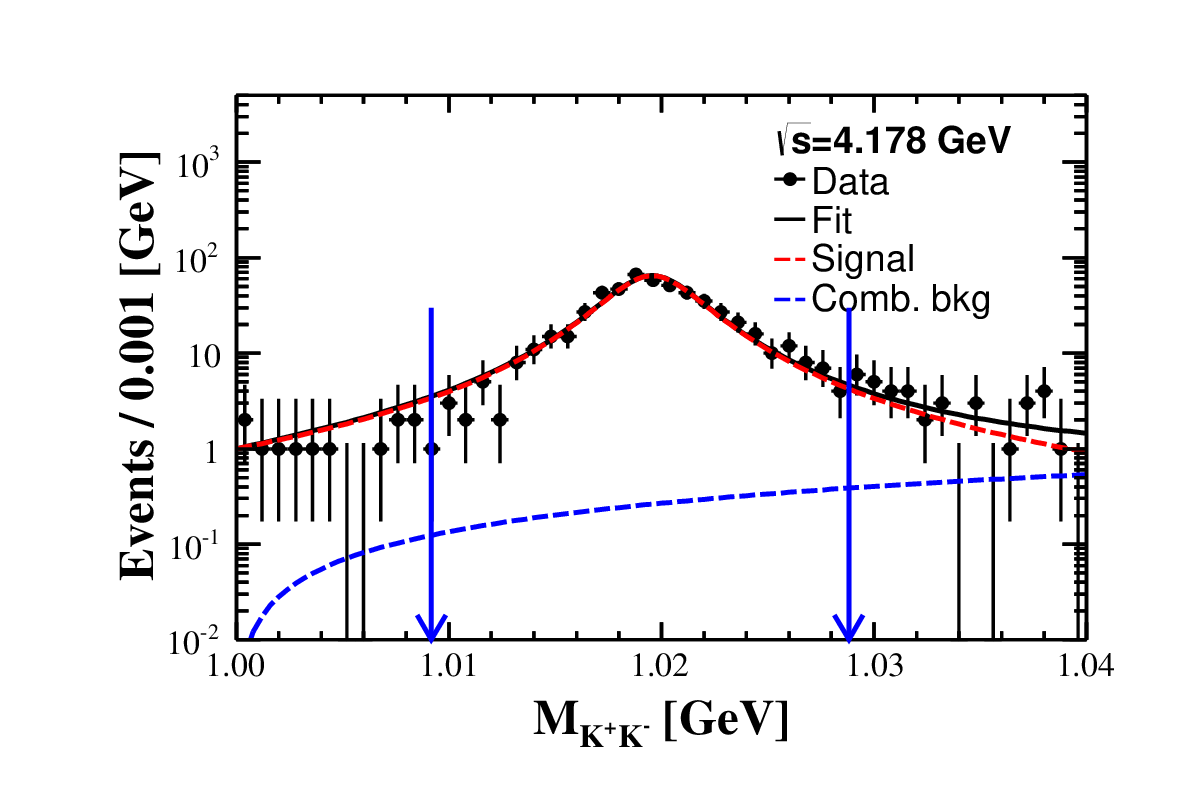}
  \setlength{\abovecaptionskip}{-11.0pt}
\setlength{\belowcaptionskip}{0.0pt}
  \caption{Spectrum of $M_{\kk}$ for data taken at $\sqrt s=4.178$~GeV. The filled black circles with error bars correspond to data. The red dashed line, blue dotted line and black curve show the signal, background and total fit result, respectively. The blue arrows indicate the window that has been applied to select signal events.}
  \label{fig::1}
\end{figure}

In addition, the $K^+K^-$ invariant mass, $M_{K^+K^-}$, is required to satisfy $|M_{K^{+}K^{-}}-M_{\phi}|<2\sigma_{M_{K^{+}K^{-}}}\approx9.8$ MeV, where $M_{\phi}$ is the $\phi$ nominal mass~\cite{ParticleDataGroup:2022pth} and $\sigma_{M_{K^{+}K^{-}}}$ corresponds to the observed width of a Breit-Wigner function. This width is experimentally determined by a fit to the $M_{K^{+}K^{-}}$ spectrum of data taken at $\sqrt s=4.178$ GeV as shown in Fig.~\ref{fig::1}. The signal is described by a Breit-Wigner function and the background is represented by a first-order Chebychev function.

\begin{figure}[h]
  \centering
  \includegraphics[width= 8.7cm,height=6.0cm]{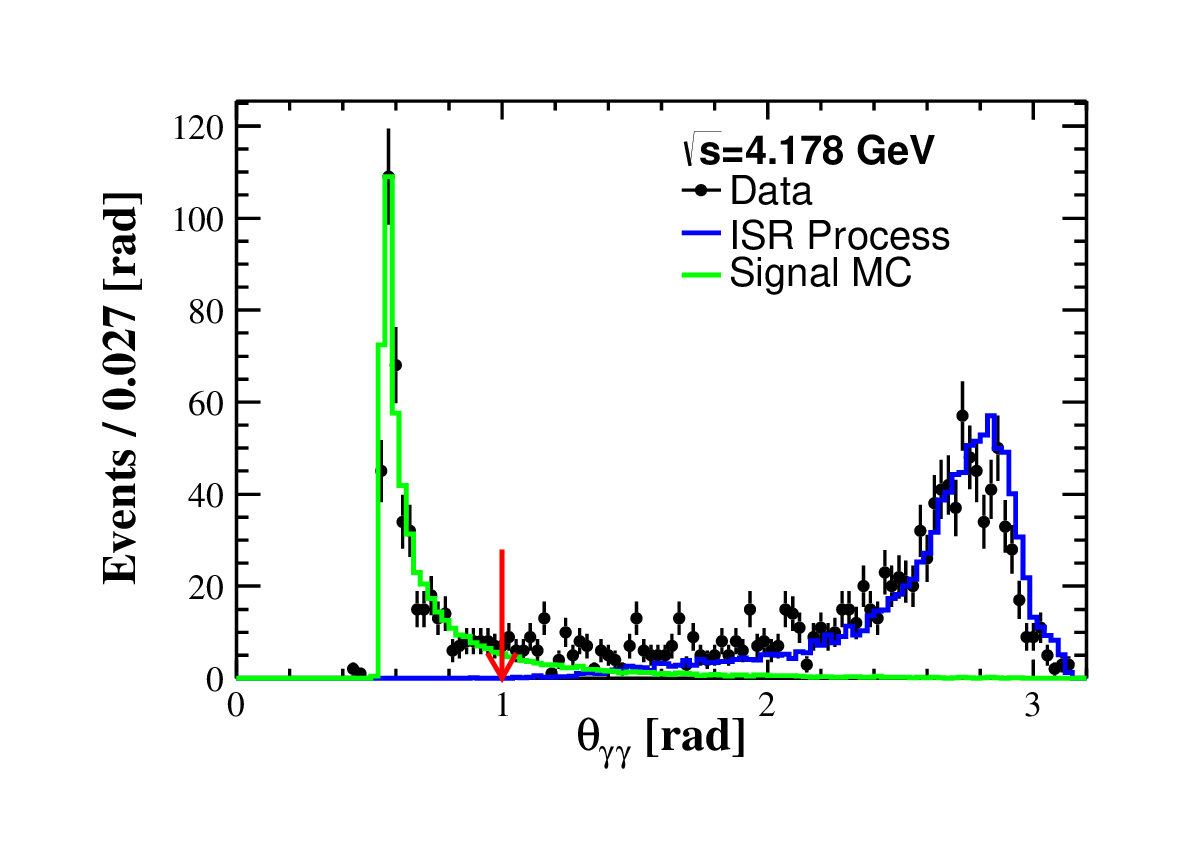}
  \setlength{\abovecaptionskip}{-11.0pt}
\setlength{\belowcaptionskip}{0.0pt}
  \caption{Distribution of the opening-angle ($\theta_{\gamma\gamma}$) of the two photon candidates for events taken at $\sqrt s=$4.178 GeV, where the black dots with error bars and the green histogram indicate data and signal MC sample, respectively, while the blue dashed histogram corresponds to the predicted contribution of the ISR process using the inclusive MC sample. The vertical red line indicates the requirement ($<1.0$ radians) that is used to select signal events.}
  \label{fig::ISR}
\end{figure}

Based on an analysis of the inclusive MC events, we find that the ISR process $\ISR$ is the dominant background, where the radiative photon is combined with fake photon candidates resulting in their invariant mass falling within the $\eta$ mass region. The fake photon candidates arise from candidates associated with the detector backgrounds, photons originating from the beam or detector interactions with other particles. To reject this type of background, we require the opening angle between the two photons in the lab frame to be less than 1.0 rad for events in which the two-photon invariant mass falls between 0.4 and 0.5~GeV. Figure~\ref{fig::ISR} illustrates the two-photon opening-angle distribution for data taken at $\sqrt s=4.178$~GeV.

\section{BORN CROSS SECTION}

\begin{figure}[h]
  \centering
  \includegraphics[width= 8.7cm,height=6.0cm]{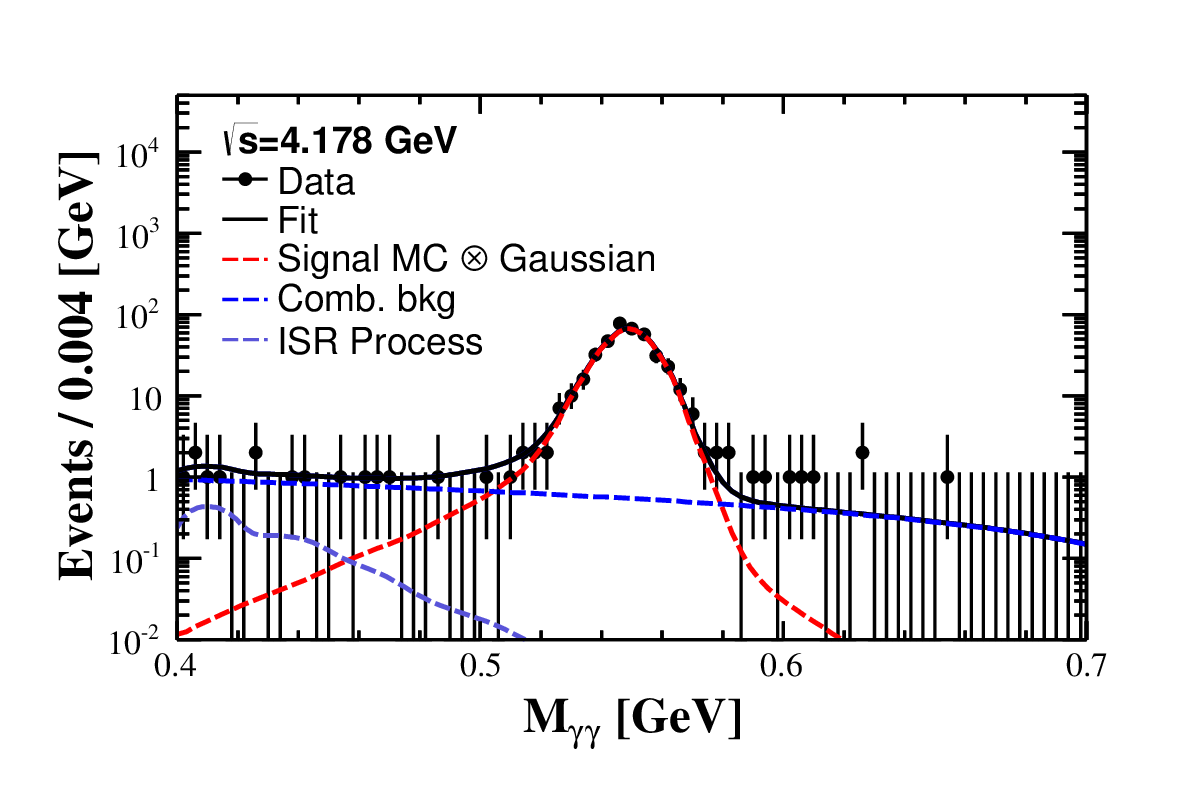}
  \setlength{\abovecaptionskip}{-11.0pt}
\setlength{\belowcaptionskip}{11.0pt}
  \caption{Results of a fit of the $M_{\gamma\gamma}$ spectrum for data taken at $\sqrt s=$4.178 GeV. The black dots with error bars are data and the black solid line shows the result of the fit including all considered contributions. The red dashed curve represents the contribution from the signal of interest. The blue dotted line and the dark purple dash-dotted line are the first-order Chebychev function and the predicted line shape obtained by analyzing the ISR process.}
  \label{fig::higlumiresult}
\end{figure}

The number of signal events is obtained by a fit to the two-photon invariant-mass ($M_{\gamma\gamma})$ spectrum with an unbinned maximum-likelihood method~\cite{Max}. The signal is described by the shape taken from signal MC convolved with a Gaussian function to account for the small difference in resolution between data and MC simulation. The mean and width of the Gaussian function are free parameters in the fit. The combinatorial background shape is described by a first-order Chebychev function, and the line shape of $\ISR$ is based on the MC study. Figure~\ref{fig::higlumiresult} shows the $M_{\gamma\gamma}$ distribution for data taken at $\sqrt s= 4.178$ GeV. The fitted signal yields for data taken at all the available energy points are summarized in Table~\ref{tabB}. Due to the limited statistics at $\sqrt s=
4.242$, 4.308,
4.527, and 4.575 GeV, upper limits at the 90\% confidence level are set at these energies, taking into account the systematic uncertainty described later.

The Born cross sections at each energy point are calculated by:
\vspace{-0.5cm}
\begin{center}
\begin{eqnarray}\label{eq:BornCrsSec}
    \sigma^{B} = \frac{N}{L_{\rm int} \cdot \varepsilon \cdot f_\text{\rm ISR}\cdot f_\text{vac}\cdot Br(\eta)\cdot Br(\phi)},
\end{eqnarray}
\end{center}
where $Br(\eta)$ and $Br(\phi)$ are the branching fractions~\cite{ParticleDataGroup:2022pth} of $\eta\to \gamma \gamma$ and $\phi\to K^+K^-$, respectively, $N$ represents the signal yield, $L_{\rm int}$ is the integrated luminosity, $\varepsilon$ is the detection efficiency determined via MC simulation.
The FSR effect, which is considered by utilizing the \textrm{PHOTOS} software~\cite{photos}, has been incorporated into the event generators.
This enables the generation of MC events that undergo detector simulation to estimate the detector efficiency.
Additionally, the ISR effect is also taken into account in this process.
The $f_\text{\rm ISR}\cdot f_\text{\rm vac}$ is the product of the ISR correction factor with the polarization factor, which is obtained by:
\vspace{-0.9cm}
\begin{center}
\begin{eqnarray}
     f_\text{\rm ISR}\cdot f_\text{\rm vac} = {1\over \sigma^{B}(s)} \int{ {\sigma^{B}(s(1-x)) \over |1+\Pi(s (1-x))|^2}F(x,s)dx}.
\end{eqnarray}
\end{center}
 Here, $\Pi(s)$ is the vacuum polarization factor~\cite{Ping:2016pms}, including leptonic and hadronic parts, $F(x,s)$ is the radiator function taken from a QED calculation~\cite{Kuraev:1985hb} with an accuracy of $0.1\%$, and $\sigma^{B}(s)$ is the Born cross section, which is taken from this analysis at $\sqrt s=3.773$ to 4.600~GeV and from BaBar~\cite{BaBar:2007ceh} and BESIII~\cite{BESIII:2021bjn} at $\sqrt s=1.560$ to 3.080~GeV. The measured Born cross sections are obtained via an iterative process till a stable result.

\begin{figure}[h]
\begin{center}
 \subfigure{\includegraphics[width=8.7cm,height=6.0cm]{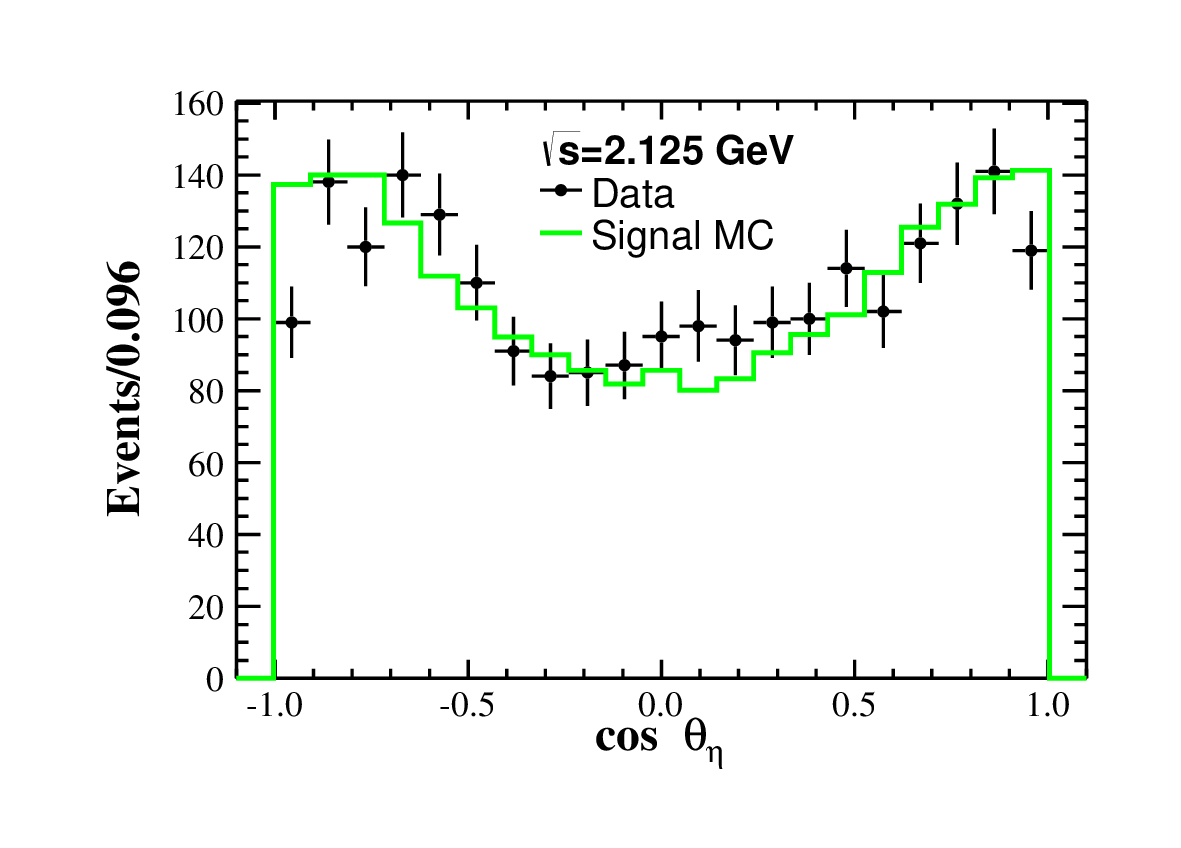}}
\setlength{\abovecaptionskip}{-9.0pt}
\setlength{\belowcaptionskip}{0.0pt}
\caption{Angular distribution of $\eta$ in the $\EE$ center-of-mass frame. The filled circles with error bars correspond to data and the solid green histogram is the signal MC simulation.}
\label{2125}
\end{center}
\end{figure}

The MC samples have been validated by comparing the modeled angular distributions of the final-state mesons with the ones extracted from data with an integrated luminosity of 108.49~$\text{pb}^{-1}$ taken at $\sqrt s=2.125$~GeV. The simulated angular distribution of the $\eta$ meson follows the expected $P$-wave dynamics.  Figure~\ref{2125} shows a comparison of the measured polar-angle distribution of the $\eta$ meson with predictions obtained using the previously described MC sample. The MC results are consistent with data within statistical uncertainties.

Table~\ref{tabB} summarizes the measured Born cross sections at the energy points in the range $\sqrt s= 3.773$ to 4.600~GeV. For most of the energy points, we observe a statistically significant ($>5\sigma$) signal of the process $\EE\to\ep$. The statistical significance of signal at each energy point is calculated according to the change of likelihood with and without the signal component
versus the change of number of degrees of freedom.

\begin{table*}[htb]
\centering
\caption{The number of $e^{+}e^{-}\to\eta\phi$ events ($N$), the integrated luminosity ($L_{\rm int}$), the detection efficiency ($\varepsilon$), the radiative correction factor with vacuum factor polarization factor ($f_{\rm ISR}\cdot f_{\rm vac}$), the Born cross section ($\sigma^{B}$) and statistical significance ($S$) at different center-of-mass energies ($\sqrt{s}$). For $\sigma^{B}$, the first and second uncertainties are statistical and systematic, respectively; for $N$, the uncertainty is statistical only. The values in brackets for $N$ and $\sigma^{B}$ correspond to the upper limits at 90\% confidence level. }
\label{tabB}
\begin{tabular}{cccccccc}
  \hline
  \hline
  $\sqrt{s}$ (GeV)& $N$ & $L_{\rm int}$ (\rm pb$^{-1}$) & $\varepsilon$ (\%)& $f_{\rm ISR}\cdot f_{\rm vac}$ &~$f_{\rm vac}$ & $\sigma^{B}(\rm pb)$ & $S$\\ \hline

   3.773    &858.5$\pm$29.8             & 2932.80 & 8.07 & 2.64  &1.057    & 7.09 $\pm$ 0.25 $\pm$ 0.24 &$>10\sigma$  \\ \hline
   4.008    &97.4$\pm$9.9               & 482.00  & 6.14 & 3.47  &1.044    & 4.89 $\pm$ 0.50 $\pm$ 0.15 &$>10\sigma$ \\ \hline
   4.086    &7.0$^{+3.2}_{-2.7}$ $(<12)$& 52.63   & 5.67 & 3.83  &1.052    & 3.15 $^{+1.44}_{-1.21}$ $\pm$ 0.10$(<5.41)$&$4.9\sigma$\\ \hline
   4.178    &393.1$\pm$20.1             & 3189.00 & 4.96 & 4.33  &1.055    & 2.96 $\pm$ 0.15 $\pm$ 0.11 &$>10\sigma$ \\ \hline
   4.189    &52.8$\pm$7.5               & 526.70  & 4.96 & 4.36  &1.056    & 2.39 $\pm$ 0.34 $\pm$ 0.08 &$>10\sigma$ \\ \hline
   4.200    &51.8$\pm$7.3               & 526.00  & 4.92 & 4.38  &1.057    & 2.36 $\pm$ 0.34 $\pm$ 0.08 &$>10\sigma$ \\ \hline
   4.210    &61.1$\pm$7.9               & 517.10  & 4.91 & 4.36  &1.057    & 2.85 $\pm$ 0.37 $\pm$ 0.09 &$>10\sigma$ \\ \hline
   4.219    &55.6$\pm$7.8               & 514.60  & 5.04 & 4.27  &1.057    & 2.59 $\pm$ 0.36 $\pm$ 0.10 &$>10\sigma$  \\ \hline
   4.226    &142.3$\pm$ 12.1            & 1091.74 & 5.20 & 4.18  &1.056    & 3.09 $\pm$ 0.26 $\pm$ 0.12 &$>10\sigma$ \\ \hline
   4.236    &59.8$\pm$8.0               & 530.30  & 5.30 & 4.11  &1.055    & 2.67 $\pm$ 0.36 $\pm$ 0.08 &$>10\sigma$ \\ \hline
   4.242    &4.9$^{+2.7}_{-2.3}$ $(<10)$& 55.59   & 5.29 & 4.11  &1.055    & 2.10 $^{+1.15}_{-0.97}$ $\pm$ 0.07$(<4.27)$&$3.3\sigma$\\ \hline
   4.244    &61.5$\pm$8.2               & 538.10  & 5.29 & 4.12  &1.054    & 2.70 $\pm$ 0.36 $\pm$ 0.08 &$>10\sigma$  \\ \hline
   4.258    &97.4$\pm$10.4              & 825.70  & 5.08 & 4.23  &1.053    & 2.83 $\pm$ 0.30 $\pm$ 0.10 &$>10\sigma$  \\ \hline
   4.267    &62.5$\pm$8.3               & 531.10  & 5.03 & 4.31  &1.053    & 2.80 $\pm$ 0.37 $\pm$ 0.11 &$>10\sigma$  \\ \hline
   4.278    &18.9$\pm$6.8               & 175.70  & 4.87 & 4.39  &1.053    & 2.59 $\pm$ 0.94  $\pm$ 0.09 &$7.3\sigma$ \\  \hline
   4.308    &4.0$^{+2.5}_{-2.1}$ $(<9)$ & 44.90   & 4.76 & 4.48  &1.051    & 2.13 $^{+1.35}_{-1.12}$  $\pm$ 0.07$(<4.85)$&$3.2\sigma$\\  \hline
   4.358    &55.7$\pm$7.8               & 540.00  & 5.11 & 4.22  &1.051    & 2.47 $\pm$ 0.35 $\pm$ 0.09 &$>10\sigma$  \\ \hline
   4.387    &6.0$^{+2.7}_{-2.5}$ $(<11)$& 55.18   & 4.89 & 4.42  &1.053    & 2.57 $^{+1.17}_{-1.07}$$\pm$ 0.09$(<4.76)$&$4.1\sigma$\\ \hline
   4.416    &93.9$\pm$10.2              & 1073.57 & 4.30 & 4.98  &1.055    & 2.11 $\pm$ 0.23 $\pm$ 0.07 &$>10\sigma$  \\ \hline
   4.467    &9.0$^{+3.2}_{-3.0}$        & 109.94  & 3.64 & 5.85  &1.055    & 1.97 $^{+0.70}_{-0.65}$ $\pm$ 0.07&$5.6\sigma$  \\ \hline
   4.527    &4.4$^{+2.5}_{-2.1}$ $(<11)$& 109.98  & 3.18 & 6.68  &1.055    & 0.96 $^{+0.55}_{-0.45}$ $\pm$ 0.04$(<2.43)$&$2.1\sigma$\\ \hline
   4.575    &2.0$^{+1.9}_{-1.5}$ $(<6)$ & 47.67   & 2.95 & 7.21  &1.055    & 1.00 $^{+0.97}_{-0.75}$ $\pm$ 0.03$(<3.05)$&$1.5\sigma$\\ \hline
   4.600    &26.7$\pm$5.6               & 566.90  & 2.80 & 7.52  &1.055    & 1.15 $\pm$ 0.24 $\pm$ 0.05 &$8.8\sigma$ \\  \hline
   \hline
 \end{tabular}
\end{table*}

\section{Systematic uncertainties}
Systematic uncertainties on the obtained cross sections mainly come from the photon detection efficiency, tracking and PID efficiency, kinematic fit, $\phi$ mass window, $\theta_{\gamma\gamma}$ window, fit method, integrated luminosity, radiative correction factors, and branching ratios of intermediate decays. In the following, we describe each of those items separately.

{\it Photon detection efficiency}: The systematic uncertainty in the reconstruction efficiency per photon is based on studies using the control sample $J/\psi \to \rho^{0}\pi^{0}$ with $\rho\to\pi^{+}\pi^{-}$ and $\pi^{0}\to\gamma\gamma$~\cite{BESIII:2017nty}.  It is estimated to be 1.0\%.

{\it Tracking and PID efficiency for kaons}: The uncertainties of the tracking and PID efficiency for kaons are studied using the control sample $\jpsi\to K^{*}(892)^0(\to K^+\pi^-) K^0_S(\to\pi^+\pi^-)+c.c.$. The efficiencies of data ($\epsilon_\text{data}$) and MC events ($\epsilon_\text{MC}$) are provided as a function of the transverse momentum and polar angle, thereby, represented in a two-dimensional matrix. The difference in the efficiencies between $\epsilon_{\rm data}$ and $\epsilon_{\rm MC}$ is calculated as
\begin{equation}
\delta_i = 1- {\epsilon_\text{MC}^i \over \epsilon_\text{data} ^i},
\end{equation}
where $i$ refers to a bin in the two-dimensional matrix. The systematic uncertainty caused by uncertainties from tracking or PID is estimated with
\begin{equation}
\delta = \sum_{i} R_i \delta_i,
\end{equation}
where $R_i$ is the fraction of the number of events in the $i$-th bin, and the sum runs over all bins containing data.

{\it Kinematic fit}: The systematic uncertainty due to the 4C kinematic fit is estimated by correcting the helix parameters of charged tracks in MC to match the data following the method described in Ref.~\cite{BESIII:2012mpj}. The difference in the detection efficiencies with and without the correction to the MC samples is taken as the uncertainty.

{\it $\phi$ mass window}: The uncertainty due to the $\phi$ mass window is estimated via a study of the variation of the efficiency depending on the applied $K^+K^-$ mass window. Here the efficiency is defined as the number of events in the $M_{K^+K^-}$ window ($|M_{K^{+}K^{-}}-M_{\phi}|<9.8$~MeV) over the number of events in the wider region 1.0$<M_{K^+K^-}<1.04$~GeV. The difference of the accepted efficiencies between data and MC simulation is determined to be about 0.2\%, which is taken as a systematic uncertainty.

{\it $\theta_{\gamma\gamma}$ selection}: The uncertainty caused by applying the $\theta_{\gamma\gamma}$ selection is estimated by studying the corresponding efficiencies in data and MC simulation with a control sample of $J/\psi\to\phi\eta$. We obtain selection efficiencies by taking the ratio of the number of events applying the condition $\theta_{\gamma\gamma}<1.0$~rad with the number of events without this requirement for both data and MC simulation. The difference in efficiencies between data and MC simulation, 0.3\%, is taken as a systematic uncertainty.

{\it Background shape}: Uncertainties due to the background shape are estimated by changing the background shape from a first-order Chebychev function and the shape of the $\ISR$ process to a second-order Chebychev function. The differences in the extracted Born cross sections between the nominal and modified procedures are taken as a systematic uncertainty.

{\it Fit range}: The uncertainty due to the fit range for the $M_{\GG}$ distribution is estimated by changing the fit range from $[0.40, 0.70]$~GeV to $[0.45, 0.65]$~GeV. The difference in the Born cross sections is taken as uncertainty.

{\it Integrated luminosity}: The integrated luminosity is measured with Bhabha events. The corresponding systematic uncertainty is determined to be 1.0$\%$~\cite{Ablikim:2013ntc,BESIII:2020oph}.

{\it Radiative correction factor}: The radiative correction factor, $f_{\rm ISR}\cdot f_{\rm vac}$, is estimated via an iterative procedure as described above. The difference between the last two iterations is taken as a systematic uncertainty.

{\it Branching fraction of intermediate state decays}: The uncertainties of the branching fractions $Br(\phi)=(49.20\pm0.50)\%$ and $Br(\eta)=(39.41\pm0.20)\%$ are taken from the PDG~\cite{ParticleDataGroup:2022pth}. The total uncertainty on the product of the two branching fractions, 1.1\%, is taken as a systematic uncertainty.

Correlated systematic uncertainties are marked with~*, and the others are independent systematic uncertainties, the total systematic uncertainty in the cross-section measurement is obtained by adding them in quadrature. Table~\ref{tab::SummarySystemU1} summarizes all the systematic uncertainties in the cross-section measurements.
 \begin{table*}[htb]
\centering
\caption{ Relative systematic uncertainties (in \%) of the measurement of $\sigma^{\rm B}$ from photon reconstruction ($\it PR$), tracking efficiency ($\it TE$), PID efficiency ($\it PID$), kinematic fit ($\it KF$), $\phi$ mass window ($\it \phi$$MW$), $\it \theta_{\gamma\gamma}$ window ($\it \theta_{\gamma\gamma}$), background shape ($\it BS$), fit range ($\it FR$), integrated luminosity ($\it IL$), radiative correction factor ($\it RCF$) and branching fraction of intermediate states decay ($\it BF$). Correlated systematic uncertainties are marked with *, and the others are independent systematic uncertainties. The total uncertainty is obtained by adding all items in quadrature. Due to the limited statistics at $\sqrt s=$ 4.086, 4.242, 4.308, 4.387, 4.527 and 4.575 GeV, the systematic uncertainties assigned due to background shape or fit range is ignored and expressed as ...~.}
\label{tab::SummarySystemU1}
\begin{tabular}{cccccccccccccccc}
  \hline
  \hline
  $\sqrt s$ (GeV) & $\it PR^*$    & $\it TE^*$    & $\it PID^*$   & $\it KF^*$    & $\it \phi$$MW^*$ & $\it \theta_{\gamma\gamma}^*$ & $\it BS$   & $\it FR$   & $\it IL^*$   & $\it RCF$  & $\it BF^*$   & Total \\ \hline
3.773&	2.0&	0.4&	1.2&	0.1&	0.2&	0.3&	0.7&	1.8&	1.0&	0.3&	1.1&	3.5\\ \hline
4.008&	2.0&	0.2&	1.1&	0.1&	0.2&	0.3&	0.2&	1.0&	1.0&	1.0&	1.1&	3.1\\ \hline
4.086&	2.0&	0.2&	1.5&	0.1&	0.2&	0.3&	... &   ...	&   1.0&	1.4&	1.1&	3.3\\ \hline
4.178&	2.0&	0.2&	1.5&	0.2&	0.2&	0.3&	1.2&	1.1&	1.0&	1.8&	1.1&	3.8\\ \hline
4.189&	2.0&	0.3&	1.5&	0.1&	0.2&	0.3&	1.1&	0.1&	1.0&	1.7&	1.1&	3.5\\ \hline
4.200&	2.0&	0.4&	1.5&	0.1&	0.2&	0.3&	0.4&	1.3&	1.0&	1.4&	1.1&	3.4\\ \hline
4.210&	2.0&	0.4&	1.7&	0.1&	0.2&	0.3&	0.1&	0.7&	1.0&	0.8&	1.1&	3.1\\ \hline
4.219&	2.0&	0.3&	1.6&	0.2&	0.2&	0.3&	1.3&	2.0&	1.0&	0.1&	1.1&	3.7\\ \hline
4.226&	2.0&	0.3&	1.5&	0.2&	0.2&	0.3&	2.0&	1.2&	1.0&	0.8&	1.1&	3.8\\ \hline
4.236&	2.0&	0.3&	1.3&	0.1&	0.2&	0.3&	0.5&	0.4&	1.0&	1.0&	1.1&	3.0\\ \hline
4.242&	2.0&	0.8&	1.9&	0.1&	0.2&	0.3&	... &   ...	&   1.0&	0.8&	1.1&	3.5\\ \hline
4.244&	2.0&	0.4&	1.7&	0.1&	0.2&	0.3&	0.4&	0.4&	1.0&	0.7&	1.1&	3.1\\ \hline
4.258&	2.0&	0.3&	1.7&	0.1&	0.2&	0.3&	1.3&	0.7&	1.0&	0.5&	1.1&	3.4\\ \hline
4.267&	2.0&	0.4&	1.8&	0.1&	0.2&	0.3&	0.0&	2.3&	1.0&	1.2&	1.1&	4.0\\ \hline
4.278&	2.0&	0.4&	1.7&	0.1&	0.2&	0.3&	0.2&	0.4&	1.0&	2.0&	1.1&	3.5\\ \hline
4.308&	2.0&	0.4&	1.7&	0.1&	0.2&	0.3&	... &   ...	&   1.0&	1.7&	1.1&	3.4\\ \hline
4.358&	2.0&	0.3&	1.4&	0.1&	0.2&	0.3&	1.7&	0.3&	1.0&	2.0&	1.1&	3.9\\ \hline
4.387&	2.0&	0.2&	2.0&	0.1&	0.2&	0.3&	... &   ... &	1.0&	1.3&	1.1&	3.4\\ \hline
4.416&	2.0&	0.3&	1.8&	0.1&	0.2&	0.3&	1.0&	0.5&	1.0&	0.9&	1.1&	3.4\\ \hline
4.467&	2.0&	0.2&	1.3&	0.2&	0.2&	0.3&	0.2&	0.9&	1.0&	1.7&	1.1&	3.4\\ \hline
4.527&	2.0&	1.0&	2.2&	0.1&	0.2&	0.3&	... &   ...	&   1.0&	1.5&	1.1&	3.8\\ \hline
4.575&	2.0&	0.2&	2.0&	0.1&	0.2&	0.3&	... &   ...	&   1.0&	1.3&	1.1&	3.5\\ \hline
4.600&	2.0&	0.2&	2.0&	0.1&	0.2&	0.3&	1.1&	1.7&	1.0&	1.2&	1.1&	3.9\\ \hline
\hline
\end{tabular}
\end{table*}

\section{Cross section for $Y$(4230)/$Y$(4360)$ \to \eta\phi$}
We search for $Y$(4230)$\to\eta\phi$ and $Y$(4360)$\to\eta\phi$ by fitting the line shape of the cross section for $e^{+}e^{-}\to\eta\phi$ with the least-squares method incorporating the correlated and uncorrelated uncertainties. The measured cross sections of all data sets are described by

\begin{multline*}\label{eq::FitBornCrsSec}
\sigma^{B}(\sqrt{s}) = \bigg| {c_{0}\over s}e^{-a\sqrt{s}} + c_1e^{i\phi_1}BW_1(\sqrt{s}) \\
+ c_2e^{i\phi_2}BW_2(\sqrt{s})\bigg|^{2}{P(\sqrt{s})^3},
\end{multline*}
with
\begin{eqnarray}
BW_i(\sqrt{s})={\sqrt{12\pi\Gamma_i}}/({s - M_i^{2} + iM_i\Gamma_i}),
\end{eqnarray}
where $BW_i(\sqrt s)$ represents the Breit-Wigner function for $Y(4230)$ or $Y(4360)$, and $e^{-a\sqrt{s}}\,c_{0}/s$ is the continuum component, where $a$, $c_0$, $c_1$, and $c_2$ are parameters that need to be determined through a fit of the data. $\phi_1$ and $\phi_2$ are the relative phase angles of the two resonances, $M_i$ and $\Gamma_i(i=1,2)$ are fixed to the masses and widths of~the $Y(4230)$ or $Y(4360)$ resonances \cite{ParticleDataGroup:2022pth}, respectively, and $P(\sqrt{s})$ is the magnitude of the momentum of the $\phi$ meson in the $\EE$ center-of-mass system.

\begin{figure*}[ht]
\centering
  \subfigure{\includegraphics[width=8.7cm,height=6.3cm]{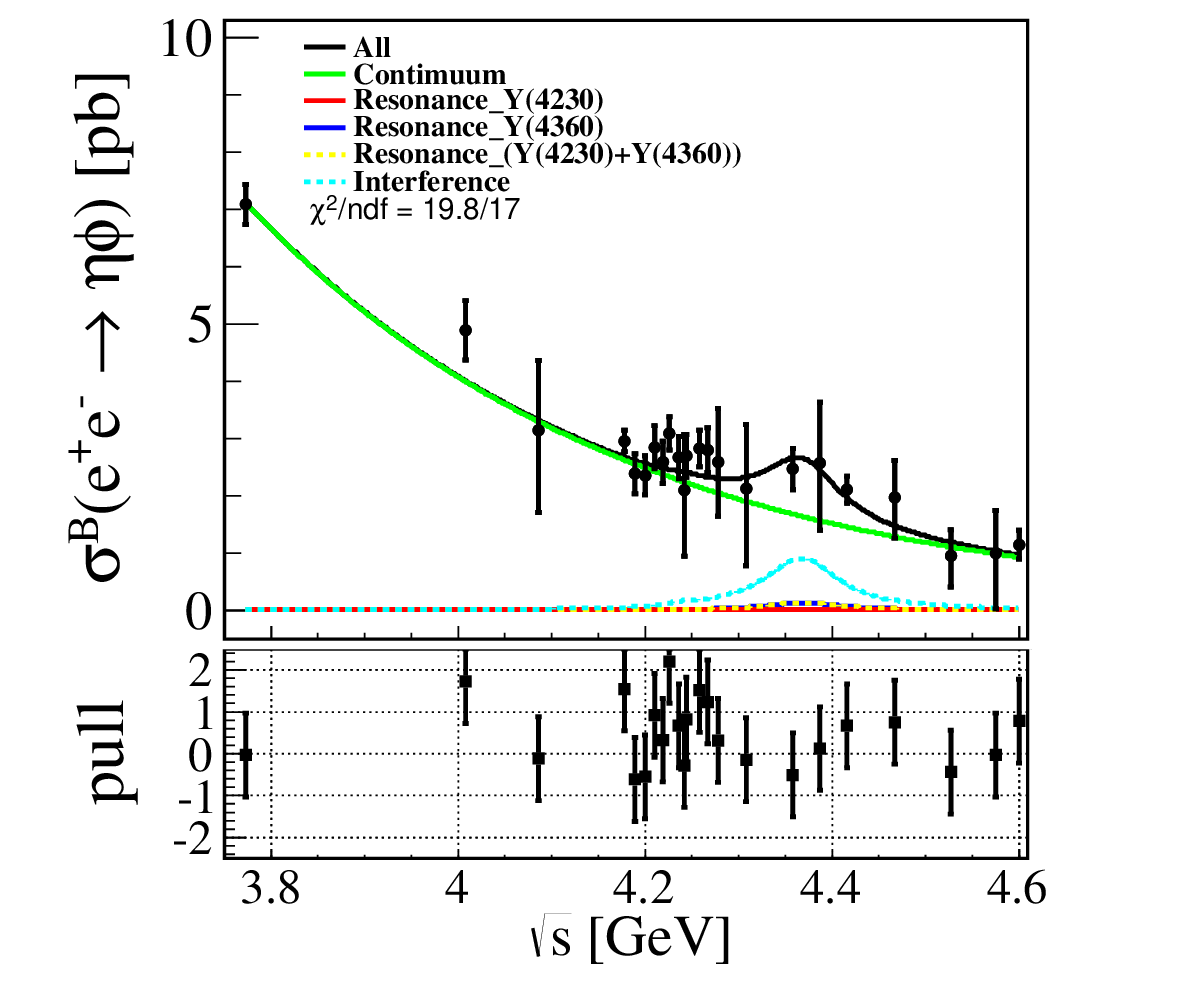}}
  \put(-70,150){\bf  ~(a)}
  \subfigure{\includegraphics[width=8.7cm,height=6.3cm]{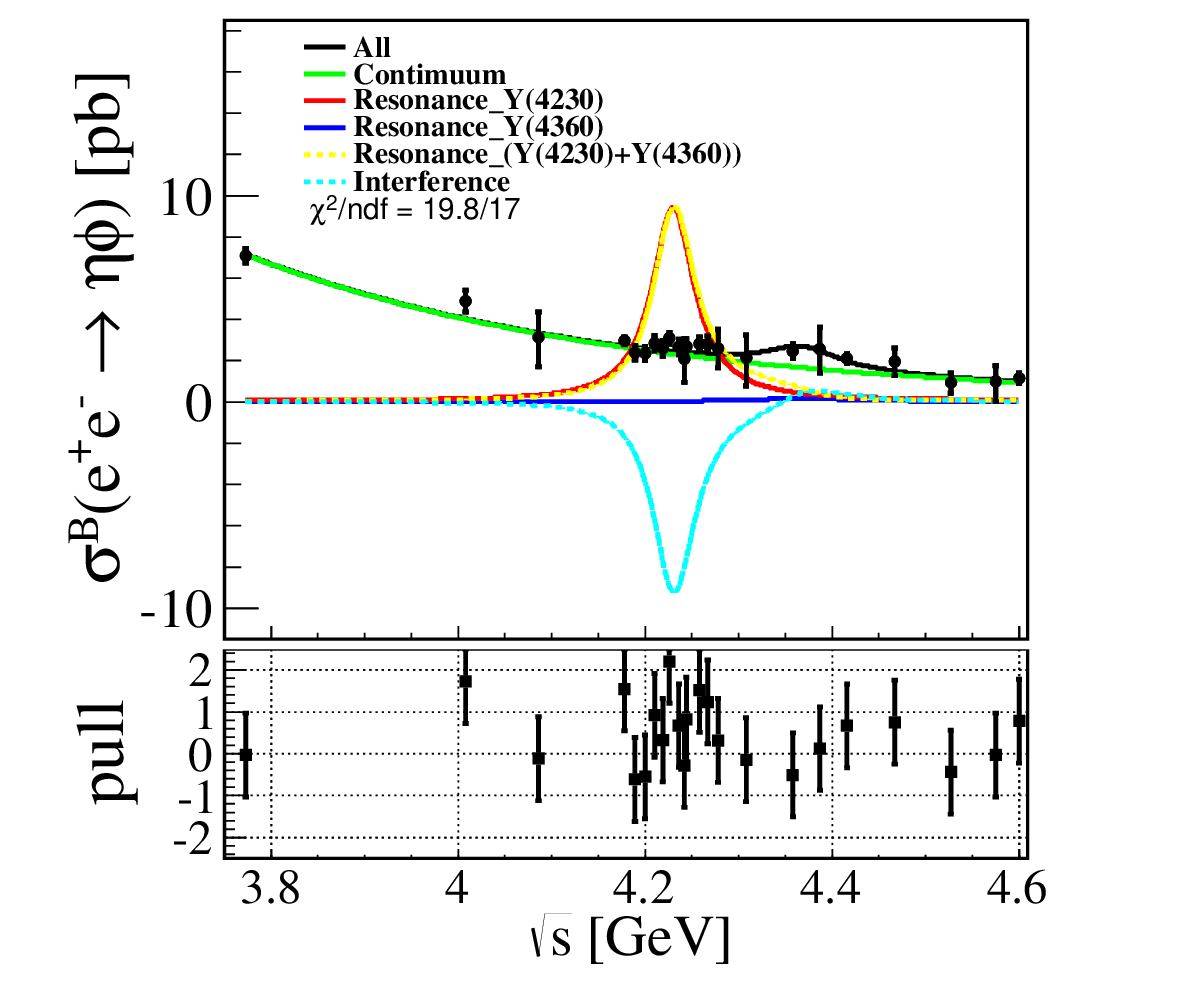}}
  \put(-70,150){\bf ~(b)}

  \subfigure{\includegraphics[width=8.7cm,height=6.3cm]{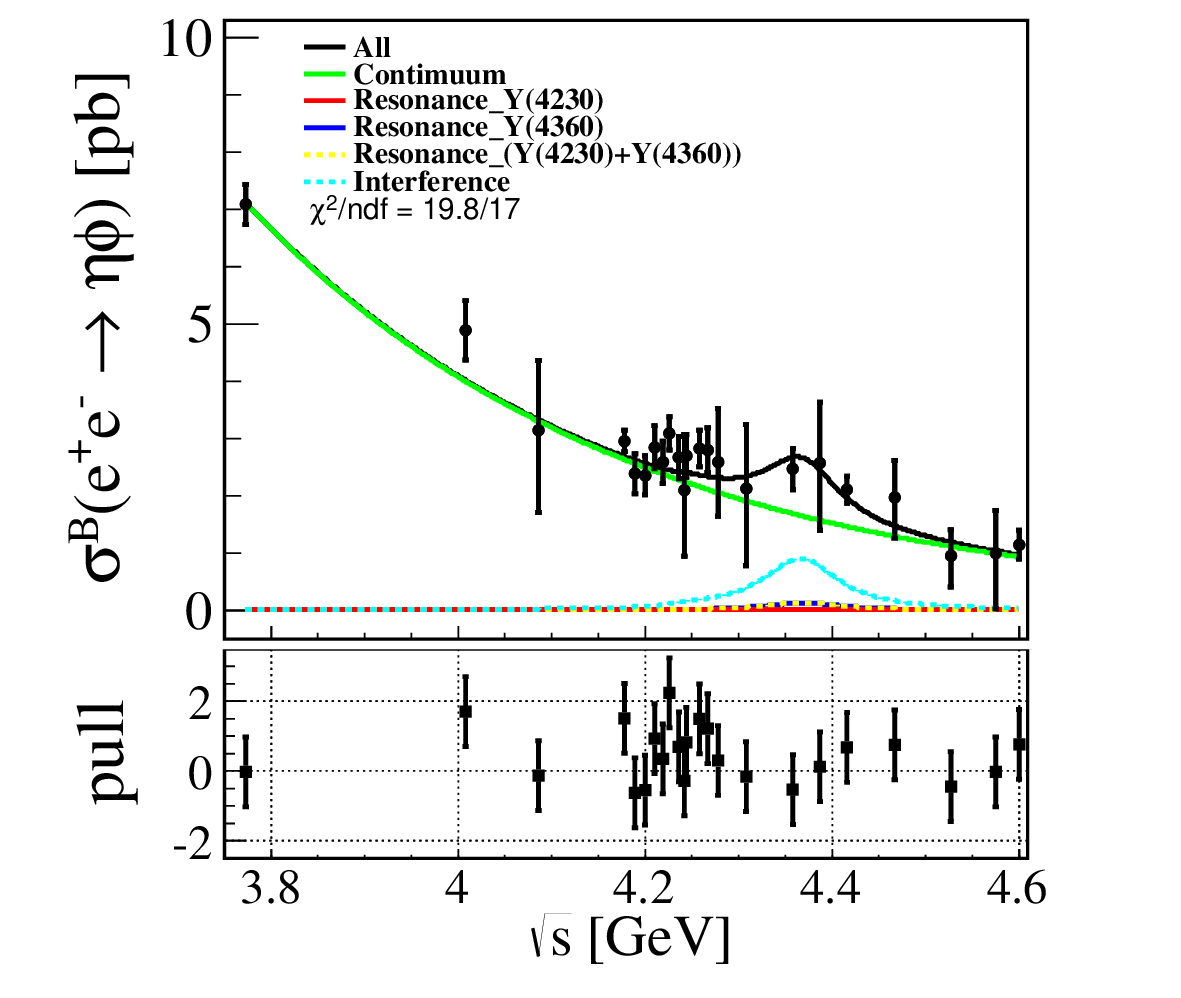}}
  \put(-70,150){\bf  ~(c)}
  \subfigure{\includegraphics[width=8.7cm,height=6.3cm]{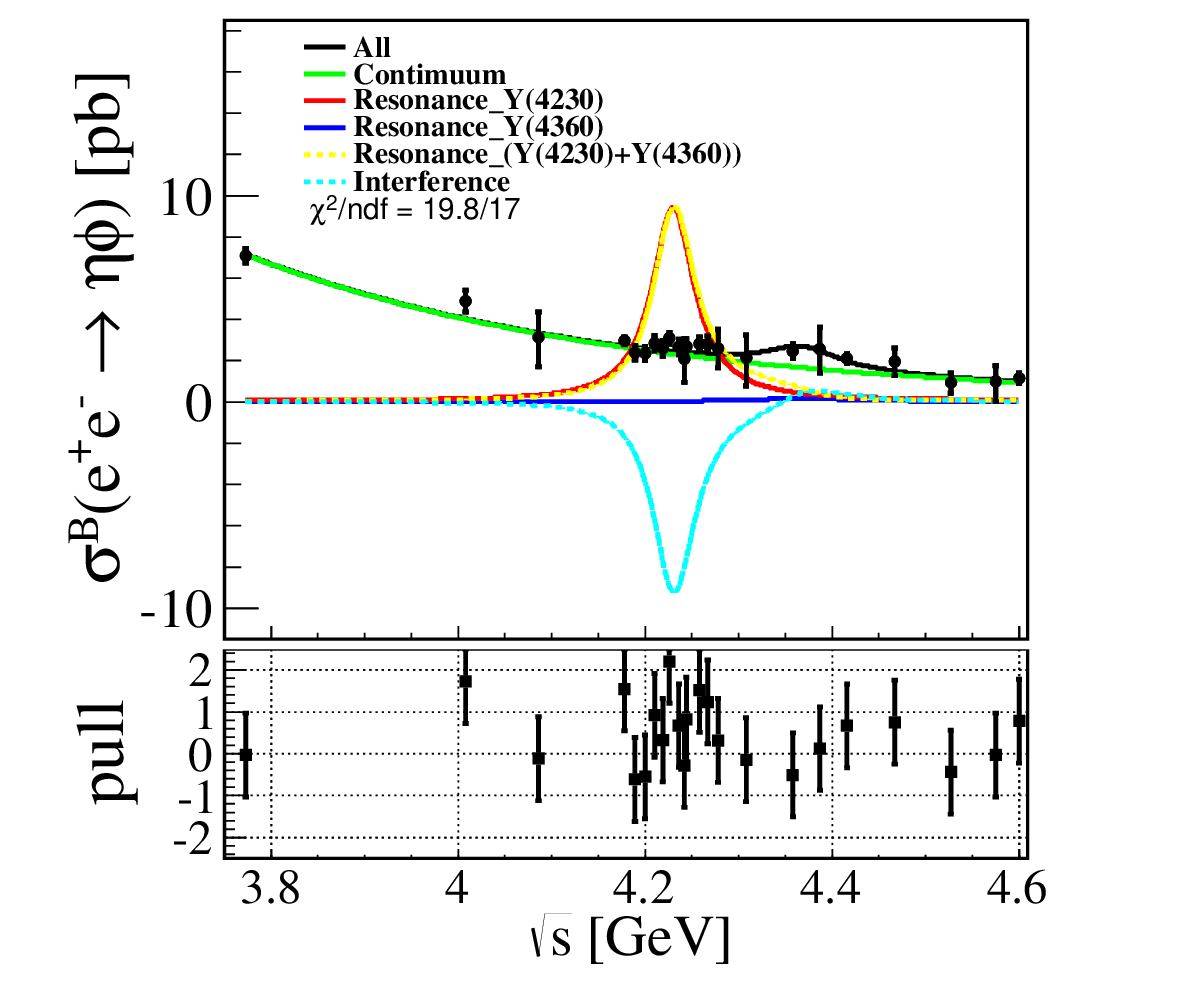}}
  \put(-70,150){\bf ~(d)}
  \setlength{\abovecaptionskip}{2.0pt}
\setlength{\belowcaptionskip}{0.0pt}
\caption{Measured Born cross sections as function of the $\sqrt{s}$. The data are subjected to a fit as described in the text. The four solutions of the fit are presented in panels (a), (b), (c) and (d). The two best and indistinguishable solutions of the fit are presented in panels (a,c) and (b,d). The black curves represent the total fit results. The individual contributions are indicated by the green curves (continuum), blue curve ($Y$(4230)), red curve ($Y$(4360)), yellow dash curve (contributions for the two resonances) and cyan dash curve (interference for the two resonances). The fit quality is presented by pull distributions and depicted in the bottom panels.}
\label{figB0}
\end{figure*}

The four solutions of the fit with equal fit quality are shown in Fig.~\ref{figB0}. The bottom panels of each plot present the pull distributions defined by $\frac{\sigma^{B}-\sigma_i}{\Delta\sigma^{B}}$, where the $\sigma^{B}$ and $\sigma_i$ are the measured and fitted cross sections at each $\sqrt{s}$, respectively, and $\Delta\sigma^{B}$ corresponds to the error of the measured value which is counted as the quadratic sum of the statistical and systematic uncertainties. For the first solution, the statistical significance of $Y(4230)$ or $Y(4360)$ is estimated to be $0.2\sigma$ or $3.2\sigma$ by comparing the changes in the $\chi^{2}$/n.d.f with or without the $Y(4230)$ or $Y(4360)$ resonance, while for other solutions, the significance remains unchanged. Here, n.d.f denotes the number of degrees of freedom. The significance for the presence of both resonances is determined to be $3.0\sigma$ for the four solutions.
We do not consider the two solutions which take into consideration a strong destructive interference (presented in Fig.~\ref{figB0}(b) and (d)) to be likely physical. The data can be described by a simplified exponential form, $e^{-a\sqrt{s}}\,c_{0}/s$, which indicates no significant contributions of the $Y(4230)$ and $Y(4360)$ states, with the fit quality $\chi^{2}$/n.d.f = 15.1/21. Moreover, the ambiguities in the fit driven by uncertainties in the phases do not allow us to provide model-independent upper limits on the coupling strengths of the $Y(4230)$ and $Y(4360)$ states in the $\eta\phi$ final state.

\section{Summary}
The Born cross sections of $\EE \rightarrow \eta\phi$ are measured with data samples corresponding to an integrated luminosity of 15.03~fb$^{-1}$ collected with the BESIII detector at 23 center-of-mass energies in the range from $\sqrt s = 3.773$ to 4.600 GeV. The line shape of the Born cross section is consistent with the continuum production, with no significant contribution from $Y(4230)$ or $Y(4360)$. To rigorously conclude on the contributions of the $Y(4230)$ and $Y(4360)$ states, it will be necessary to improve the precision of the Born cross section measurements by a significant increase of the integrated luminosity in $e^{+}e^{-}$ collisions at the respective center-of-mass energies and perform a simultaneous coupled-channel fit including various final states.

\begin{acknowledgements}
The BESIII Collaboration thanks the staff of BEPCII and the IHEP computing center for their strong support. This work is supported in part by National Key R\&D Program of China under Contracts Nos. 2020YFA0406300, 2020YFA0406400; National Natural Science Foundation of China (NSFC) under Contracts Nos. 12175244, 11875115, 12275058, U2032110, 11635010, 11735014, 11835012, 11935015, 11935016, 11935018, 11961141012, 12022510, 12025502, 12035009, 12035013, 12061131003, 12192260, 12192261, 12192262, 12192263, 12192264, 12192265, 12221005, 12235017; the Chinese Academy of Sciences (CAS) Large-Scale Scientific Facility Program; the CAS Center for Excellence in Particle Physics (CCEPP); CAS Key Research Program of Frontier Sciences under Contracts Nos. QYZDJ-SSW-SLH003, QYZDJ-SSW-SLH040; 100 Talents Program of CAS; The Institute of Nuclear and Particle Physics (INPAC) and Shanghai Key Laboratory for Particle Physics and Cosmology; ERC under Contract No. 758462; European Union's Horizon 2020 research and innovation programme under Marie Sklodowska-Curie grant agreement under Contract No. 894790; German Research Foundation DFG under Contracts Nos. 443159800, 455635585, Collaborative Research Center CRC 1044, FOR5327, GRK 2149; Istituto Nazionale di Fisica Nucleare, Italy; Ministry of Development of Turkey under Contract No. DPT2006K-120470; National Research Foundation of Korea under Contract No. NRF-2022R1A2C1092335; National Science and Technology fund of Mongolia; National Science Research and Innovation Fund (NSRF) via the Program Management Unit for Human Resources \& Institutional Development, Research and Innovation of Thailand under Contract No. B16F640076; Polish National Science Centre under Contract No. 2019/35/O/ST2/02907; The Swedish Research Council; U. S. Department of Energy under Contract No. DE-FG02-05ER41374
\end{acknowledgements}


\end{document}